\documentclass[useAMS,usenatbib]{mn2e}
\usepackage{graphicx}

\newcommand{\ra}{\mbox{$\alpha_{J2000}$}}
\newcommand{\dec}{\mbox{$\delta_{J2000}$}}

\newcommand{\dg}{\mbox{$^\circ$}}
\newcommand{\hr}{\mbox{$^{\rm h}$}}
\newcommand{\am}{\mbox{$^{\prime}$}}
\newcommand{\as}{\mbox{$^{\prime\prime}$}}

\newcommand{\kmps}{\mbox{km~s$^{-1}$}}

\newcommand{\mass}{\mbox{M}}
\newcommand{\radius}{\mbox{R}}
\newcommand{\temp}{\mbox{T}}

\newcommand{\rsun}{\mbox{\radius$_{\sun}$}}
\newcommand{\msun}{\mbox{\mass$_{\sun}$}}

\title[A New K7 Dwarf Eclipsing Binary]{A New Detached K7 Dwarf Eclipsing Binary System}
\author[Young et al.]{
  T.~B.~Young$^1$,
  M.~G.~Hidas$^{1}$\thanks{E-mail: mgh@phys.unsw.edu.au},
  J.~K.~Webb$^{1}$,
  M.~C.~B.~Ashley$^{1}$,
  \newauthor 
  J.~L.~Christiansen$^{1}$,
  A.~Derekas$^2$,
  C.~Nutto$^1$
\\
$^1$~School of Physics, University of New South Wales, Sydney, NSW, 2052, Australia\\
$^2$~School of Physics, University of Sydney, NSW, 2006, Australia}
\begin{document}

\date{\today}

\pagerange{\pageref{firstpage}--\pageref{lastpage}} \pubyear{2006}

\maketitle

\label{firstpage}

\begin{abstract}
We present an analysis of a new, detached, double-lined eclipsing binary system with K7~Ve components, discovered as part of the University of New South Wales Extrasolar Planet Search. The object is significant in that only 6 other binary systems are known with 
comparable or lower mass.  Such systems offer important tests of mass--radius
theoretical models.  Follow-up photometry and spectroscopy were obtained with the 40-inch and 2.3m telescopes at SSO respectively. An estimate of the radial velocity amplitude from spectral absorption features, combined with the orbital inclination (83.5$^\circ$) estimated from lightcurve fitting, yielded a total mass of M$_{total}=1.041\pm0.06$ \msun\ and component masses of M$_{A}=0.529\pm0.035$ \msun\ and M$_{B}=0.512\pm0.035$ \msun. The radial velocity amplitude estimated from absorption features (167$\pm$3 \kmps) was found to be less than the estimate from the H$_{\alpha}$ emission lines (175$\pm$1.5 \kmps).  The lightcurve fit produced radii of \radius$_{A}=0.641\pm0.05$ \rsun\ and \radius$_{B}=0.608\pm0.06$ \rsun, and a temperature ratio of \temp$_{B}/\temp_{A}=0.980\pm0.015$. The apparent magnitude of the binary was estimated to be $V=13.9\pm0.2$. Combined with the spectral type, this gave the distance to the binary as $169\pm14$~pc. The timing of the secondary eclipse gave a lower limit on the 
eccentricity of the binary system of $e\geq 0.0025\pm0.0005$.  This is the most statistically significant non-zero eccentricity found for such a system, possibly suggesting the presence of a third companion.
\end{abstract}



\section{Introduction}

Low-mass stars ($\mass<\msun$) make up more than 70\% of the stellar population, but their intrinsic faintness means that relatively few have been studied in detail. In recent years, much work has been done on the theoretical modelling of such stars (see review by \citealt{Chab&Baraffe}), but progress has been hindered by a lack of observed low-mass stars with sufficiently accurate parameters. In order to distinguish between the various models, uncertainties of less than 2--3\% in the masses and radii are required \citep{RibasReview,Ribas2003}. This can most easily be achieved with detached, double-lined eclipsing binary stars, with best results coming from systems with similar components \citep{RibasReview}. Currently, only six such systems with masses of 0.6~\msun\ or below are known. A number of recent studies of these systems \citep{GUBoo,Torres&Ribas,Maceroni,Ribas2003} have revealed discrepancies between the observations and the latest theoretical models, which in particular appear to underestimate the radii by as much as 10--15\%. This is possibly due to the enhanced magnetic activity produced by the high rotational velocities of stars in close binaries \citep{RibasReview}. There is a clear need for more of these systems to be discovered and studied, so that the physics of low-mass stars may finally be understood. 

In this paper we report the discovery and preliminary analysis of a new detached, double-lined eclipsing binary system with K7~Ve components. The system was detected in data obtained for the University of New South Wales (UNSW) Extrasolar Planet Search \citep{Hidas2005}. The eclipses are grazing, and their depth was diluted by light from a brighter, blended star. The resulting shallow eclipses led to this object's selection as as a transiting planet candidate, its true nature only being revealed by follow-up observations.


\section{Observations}

\subsection{Photometry}

The data leading to the detection of UNSW-TR-2 (\ra~=~18\hr~30\am~52\farcs3, \dec~=~$+7$\dg~9\am~27\as) were obtained using the 0.5~m Automated Patrol Telescope (APT) at Siding Spring Observatory (SSO), NSW, Australia. The object was found in a field centred on the open cluster NGC~6633. During three seasons (2002--2004) we observed this field on 79 nights (about 3500 images). For most of the present analysis, we have used only the best quality data (highest cadence and lowest photometric noise), consisting of 997 images from 11 nights in July and August 2002. On these nights we observed through a Johnson $V$ filter at a cadence of about 16 per hour. The remaining data on this field have one quarter the cadence (we were observing three other fields in parallel) and are affected by significant systematic trends on many nights. For the period determination (sec.~\ref{perecc}), we used the entire dataset. Lightcurves were extracted using a custom-built automated reduction pipeline. For more information about the telescope, the data and the reduction process, see \cite{Hidas2005}.

Initial inspection of the lightcurve of UNSW-TR-2 revealed eclipses with a depth of $\sim$25~mmag, a duration of 2~hours, and a binary period of $\sim2.1$~days. The out-of-eclipse magnitude is $V=12.0\pm0.1$, but inspection of a Digitised Sky Survey (DSS)\footnote{The Digitised Sky Survey was produced at the Space Telescope Science Institute under U.S. Government grant NAG W-2166, based on photographic data obtained using the Oschin Schmidt and UK Schmidt telescopes.} image revealed that the APT photometry aperture contains at least two blended stars of similar brightness. The APT has 9.4-arcsecond pixels, and in a crowded field such as this one (at galactic latitude +8\dg), such blending is common. The photometry aperture is 3 pixels in radius.

In June 2004 we obtained images of UNSW-TR-2 at higher spatial resolution in $V$ and $I$ using the 40-inch telescope at SSO. The images were sampled by 0\farcs 6 pixels. We observed one complete eclipse (in $\sim3$\as\ seeing), with a depth of 0.14~mag. The eclipse has the same depth and shape (within the photometric errors) in the two colour bands, suggesting that the two stars are of similar temperature.

The binary system is in fact the second-brightest object in the APT photometry aperture. It is $1.48\pm0.03$~mag fainter than the brightest one, and separated from it by $\sim$10\arcsec. The Tycho-2 catalogue \citep{Hog2000} gives a (Johnson) $V$ magnitude of $12.4\pm0.2$ for this bright neighbour. The combined apparent magnitude for the binary system is therefore $V=13.9\pm0.2$.

\begin{table}  
\begin{center}
\caption{Infrared magnitudes of UNSW-TR-2 and its close neighbour from the 2MASS catalogue.}
\label{catdata}
\begin{tabular}{lcc} 
\hline\hline
    & TR-2 & Neighbour \\
\hline
$J$ & 10.9 & 10.7 \\
$H$ & 10.2 & 10.2 \\
$K$ & 10.1 & 10.1 \\
\hline\hline
\end{tabular}
\end{center}
\end{table}

\subsection{Spectroscopy}

\subsubsection{Spectra Obtained}
\label{spectroscopy}
\begin{figure*}
\begin{center}
\includegraphics[angle=-90,width=15cm]{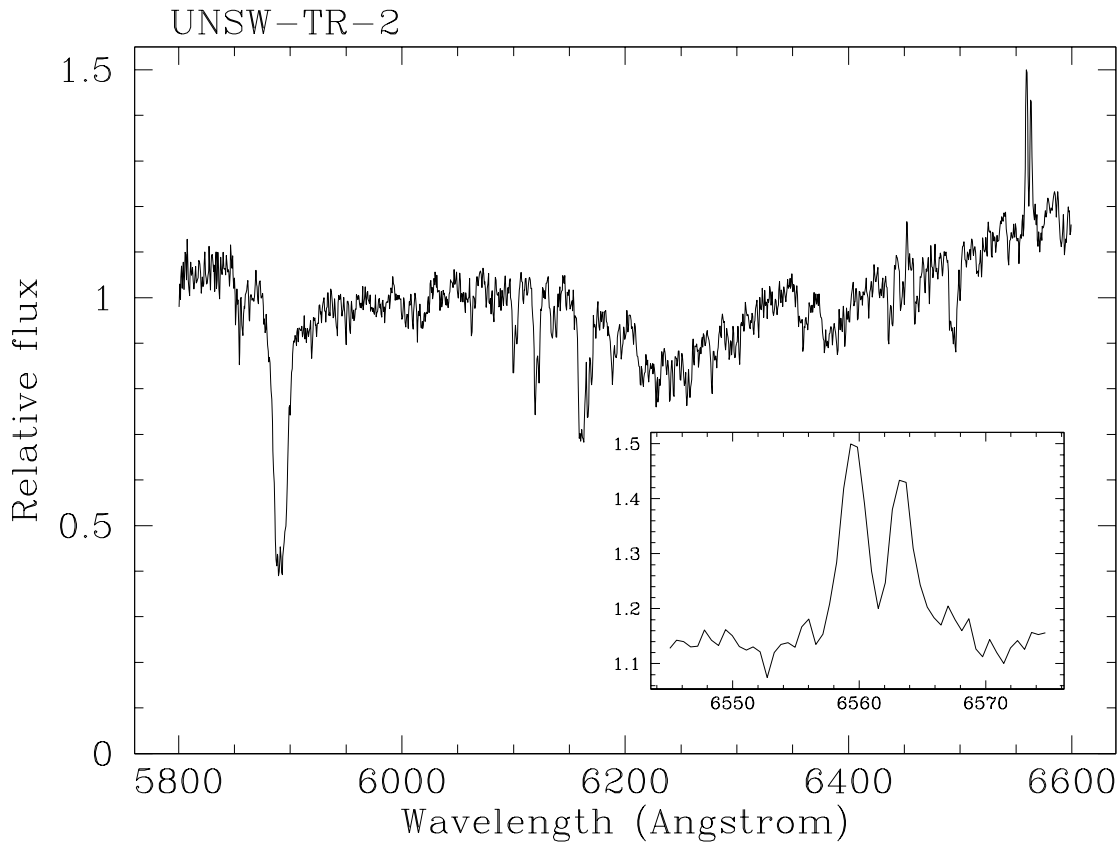}
\caption{\small The spectrum of UNSW-TR-2, obtained at phase 0.213 using the 2.3~m telescope at SSO. The inset shows a close-up of the H$_{\alpha}$ lines.}
\label{spectrum}
\end{center}
\end{figure*}

Spectra were obtained using the Double-Beam Spectrograph (DBS) on the 2.3~m telescope at SSO. During our initial follow-up run in June 2004, we obtained two spectra of UNSW-TR-2 at the same epoch (Figure~\ref{spectrum}). These covered the range 5800--6600~\AA\ with 60~\kmps\ resolution (FWHM, as measured from the sky lines in the spectra).

Visual comparison of the flux-calibrated spectra with templates from the UVILIB spectral library \citep{Pickles1998} suggested that both stars are of type K7~V. With additional spectra reaching further into the red, our estimate of the spectral type could be refined using the TiO absorption bands \citep[e.g.][]{Kenyon1987,Schild1992}. One TiO band is present in our existing data, but the absorption is very weak.

All of the absorption lines have two components, the smaller of which has 60--80\% of the depth of the larger component, pointing to two similar stars. Prominent H$_{\alpha}$ emission is also present, with the two components at a similar strength-ratio to the absorption-line pairs (Fig.~\ref{spectrum}).

The expected absolute magnitude of the two stars combined, based on their spectral type (K7~V), is $M_V=7.6$ \citep{Cox2000}. Combining this with our estimate of the system's apparent magnitude ($V=13.9\pm0.2$), and assuming 1~mag~kpc$^{-1}$ of extinction, puts the binary at a distance of $169\pm14$~pc. The template best matching the spectrum of the bright neighbouring star is that of a G8III giant, with $M_V=0.8$. Assuming the same extinction gives a distance of $1200\pm70$~pc. Therefore this neighbour is not physically associated with the binary system.

\begin{figure*}
\begin{center}
\includegraphics[bb=70 360 550 720, width=17cm,angle=0]{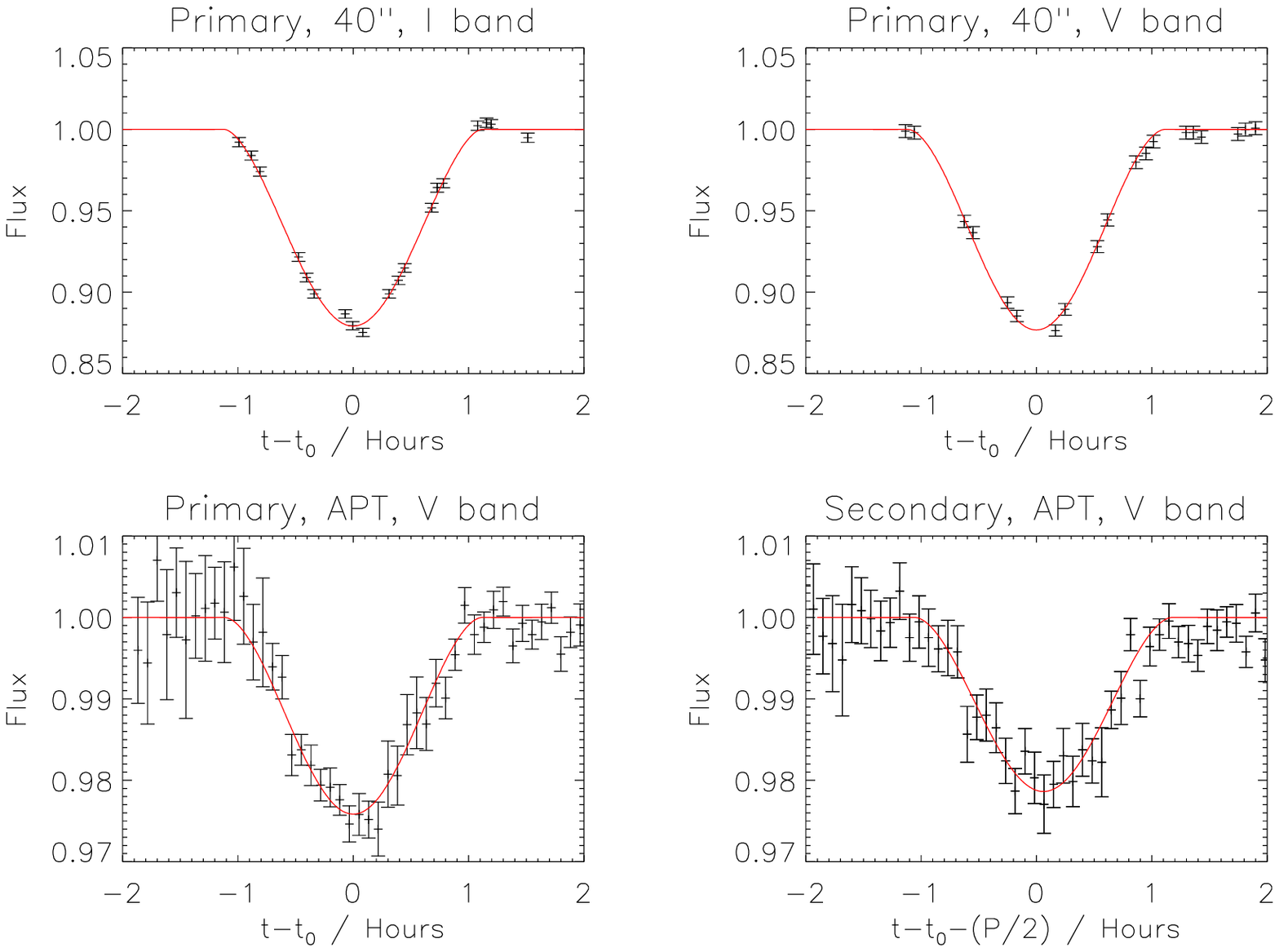}
\caption{\small The four eclipses used in the parameter estimation. Top: The primary eclipse observed in both $V$ and $I$ with the ANU 40-inch telescope. Bottom: Both eclipses from the APT lightcurve, phase folded and binned at 5-minute intervals. The solid line shows the best-fit lightcurve.}
\label{lightcurve}
\end{center}
\end{figure*}

\subsubsection{Radial Velocity Estimation}\label{rvest}
We fitted for the two stellar components simultaneously using VPFIT\footnote{http://www.ast.cam.ac.uk/$\sim$rfc/vpfit.html}. For both spectra, five absorption-line pairs with high signal-to-noise were individually fitted with blended Voigt profiles. When given the instrumental resolution, VPFIT fits for the central velocity, column density and Doppler broadening of each line and returns associated uncertainties based on the covariance matrix at the best-fit. Each absorption-line pair is surrounded in our spectra by weak absorption features, some of which are partially blended with the pair in question. To allow for this, we inserted extra absorption lines into VPFIT as appropriate, again allowing all line-parameters to vary. We found that in order to obtain reasonable fits it was also necessary to allow for the presence of a final weak subsidiary line internal to the main absorption-line pair. In all cases, this procedure produced a reasonable fit to the data ($\chi^{2}\sim1$). 

A small adaptation of VPFIT allows for the fitting of emission lines with blended Gaussian profiles. In this way we fitted the two H$_{\alpha}$ emission-line-pairs. No subsidiary lines were required to obtain a good fit.

Taking orbital phase into account, and assuming a circular orbit, the total velocity-amplitude $(K_A+K_B)\sin i$ is $167\pm3$~\kmps\ when measured from the absorption features, and $175\pm1.5$~\kmps\ when measured from the H$_{\alpha}$ emission lines, where $i$ is the orbital inclination of the system (subsequently determined by lightcurve fitting, as described in section~\ref{lcfit}). An apparent disagreement between velocities derived by these two methods has been noted in other systems (M. L{\'o}pez-Morales, private communication). The absorption velocity is considered to be more reliable due to uncertainties as to the spatial distribution of the source of H$_{\alpha}$ emission relative to the photometric centre of each star. We therefore adopt the value of $(K_A+K_B)\sin i =167\pm3$~\kmps\ for the velocity amplitude of our system. We were unable to improve upon this estimate using the two-dimensional cross-correlation software TODCOR (\cite{TODCOR}) with a variety of different template spectra. This is possibly due to the relatively low resolution of our spectra when compared to those of similar objects for which TODCOR has been successfully implemented \cite[e.g.][]{GUBoo}.

In March 2006 we obtained an additional spectrum with the same instrument and at the same resolution. This one was taken close to an eclipse, therefore the lines are completely blended and the separation of the absorption lines cannot be reliably measured. However, comparing the central wavelength of each blended line with the wavelengths of the separated components in the original spectra allows us to constrain the difference between the two masses. This was done for the five high-signal-to-noise absorption lines in both of the original spectra, leading to a constraint of
 $\Delta \mass=\mass_A-\mass_B=(0.016\pm0.03)\mass_{total}$.


\section{Analysis}

\subsection{Period and Eccentricity}\label{perecc}
Phase-folding the entire APT lightcurve (with observed eclipses up to two years apart) allows a precise determination of the orbital period. The period giving the smallest scatter in the folded lightcurve is $2.11674\pm0.00002$ days.

The epoch of the primary eclipse minimum (t$_0$) was determined from the 40-inch lightcurve. The secondary eclipse occurs at a phase of 0.5016$\pm$0.0003 (measured from the APT lightcurve), $5\pm1$~minutes later than expected for a perfectly circular orbit. Assuming that the major axis of the orbit lies in the plane of the sky gives a lower limit on the orbit's eccentricity of $e\geq 0.0025\pm0.0005$. \cite{circular} predict that a binary consisting of two 0.5~\msun\ stars with a period of about 7 days and an initial eccentricity of 0.3 will have been circularised by tidal interactions to $e\simeq0.005$ by an age of 10$^6$ years, and will have reached $e\simeq0.004$ after 10$^{10}$ years. However, the shorter period of our system should lead to increased circularisation and lower $e$. Although our lower limit may be compatible with a simple, main sequence binary system, it suggests the possibility that our system either is very young, or more likely, has at least one other companion affecting the timing of the eclipses. This is the most significant evidence for non-zero eccentricity in a low-mass eclipsing binary system detected to date. However, we continue to assume $e=0$ for this work.

\subsection{Lightcurve Fitting}\label{lcfit}
We fitted models (based on equations by \cite{Sackett1999}, modified to apply to binary stars) to the lightcurves using a chi-squared minimisation algorithm to estimate properties of the system. The APT lightcurve was folded and binned at 5-minute intervals. All the data (one high-resolution primary eclipse in each of $V$-band and $I$-band from the 40-inch, and both eclipses in $V$ from the APT) were fitted simultaneously. The accuracy of our results is limited by the lack of high-spatial-resolution observations of the secondary eclipse, and also by the presence of the nearby brighter star in the APT aperture. The epochs of the primary and secondary minima, the period, and the velocity amplitude were fixed at their previously measured values whilst allowing the orbital inclination ($i$), the radii of both stars ($\radius_A$ and $\radius_B$), and their temperature ratio ($\temp_{B}/\temp_{A}$) to vary. Also required is the fraction $F$ of the total flux in the APT aperture produced by our object, estimated as $0.19\pm0.02$ from the 40-inch images and allowed to vary within this error-range. Finally, we assume solar-metallicity limb-darkening coefficients of 0.9 and 0.7 in $V$ and $I$ respectively \citep{Claret1998}. The best-fit lightcurve is shown in Figure~\ref{lightcurve}, and the parameters are summarised in Table~\ref{parameter_table}.

\subsection{Error Estimation}
\begin{figure}
\begin{center}
\includegraphics[bb=80 380 565 700, width=9cm,angle=0]{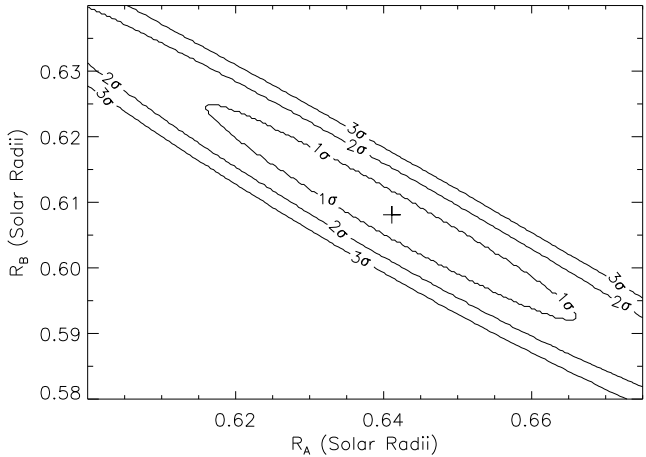}
\caption{\small The \radius$_A$ vs. \radius$_B$ chi-squared contour plot around the best-fit solution to the lightcurve. The other parameters are set to their best-fit values. The three contours represent the 1$\sigma$, 2$\sigma$ and 3$\sigma$ confidence regions, but to estimate the true uncertainties $\temp_{B}/\temp_{A}$ and $i$ must also be varied.}
\label{contours}
\end{center}
\end{figure}

To estimate the uncertainties in our lightcurve fit we plotted $\radius_A$ vs. $\radius_B$ reduced-chi-squared contour plots for ranges of $i$, $\temp_{B}/\temp_{A}$ and $F$ either-side of the best-fit parameters (Fig.~\ref{contours}). The contours on each $\radius_A$ vs. $\radius_B$ plot are ellipses, tightly constraining the sum of the radii but not their difference. Altering the other parameters from their best-fit values decreases the size of the contours whilst moving their centre. Increasing $i$ leads to a higher $\radius_A$ but lower $\radius_B$, whilst increasing $\temp_{B}/\temp_{A}$ has the opposite effect: lower $\radius_A$ and higher $\radius_B$. We found that the flux ratio $F$ is constrained relatively tightly and therefore contributes little to the uncertainties of the other parameters. Having scaled the errors such that the best-fit reduced-chi-squared has a value of one, the $1\sigma$ errors for the remaining four parameters were estimated by the range of the appropriate chi-squared contour over the full 4-dimensional error ellipsoid. Further (minor) contributions to these errors come from the uncertainties in the velocity amplitude and limb-darkening coefficients. Because the estimates of the two radii are not independent, we also used the contour plots to derive an estimate of $\Delta \radius=\radius_{A}-\radius_{B}$.

\begin{figure}
\begin{center}
\includegraphics[bb=80 375 575 700,width=9cm,angle=0]{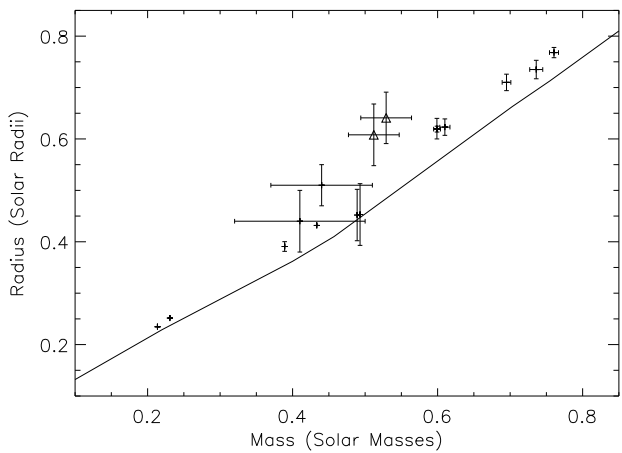}
\caption{\small Mass vs. radius for all known components of double-lined eclipsing binary stars with $M<0.8$~\msun. UNSW-TR-2 is shown as triangles. The solid line is an approximation to a 300-Myr isochrone calculated with models from \citet{Baraffe1998}. The figure is based on Fig.~1 in \citet{RibasReview}. }
\label{binaries}
\end{center}
\end{figure}

\begin{table}  
\begin{center}
\caption{Physical properties of UNSW-TR-2, derived from radial-velocity measurements (sec.~\ref{rvest}) and lightcurve fitting (sec.~\ref{lcfit}).}
\label{parameter_table}
\begin{tabular}{cr@{ $\pm$ }l} \hline\hline
Period (days)&2.11674&0.00002\\
t$_0$ (HJD) &2453171.0371&0.0005 \\
Separation (AU) &0.0327 & 0.0007\\
Inclination (degrees) &83.49 & 0.2\\
M$_{total}$ (\msun) & 1.041 & 0.06\\
 \hline
M$_A$ (\msun)  &0.529& 0.035\\
M$_B$ (\msun) &0.512& 0.035\\
$\Delta$M (\msun)  &0.017& 0.03\\
\radius$_A$ (\rsun) &0.641& 0.05\\
\radius$_B$ (\rsun) &0.608& 0.06\\
$\Delta$\radius\ (\rsun)  &0.033& 0.06\\
\temp$_B$/\temp$_A$ & 0.980&0.015\\
\hline\hline
\end{tabular}
\end{center}
\end{table}

\section{Results and Discussion}

The parameters we have derived for UNSW-TR-2 from both the lightcurve fitting and spectral analysis are shown in Table~\ref{parameter_table}. The derived masses are consistent with our earlier estimate (based on spectral shape) of two K7~V stars. The star estimated to have a higher mass, radius and temperature corresponds to the component identified in section~\ref{spectroscopy} as having stronger absorption and emission features.

In Figure~\ref{binaries} we plot the masses and radii of all the known components of detached, double-line eclipsing binaries with masses below 0.8~\msun, including UNSW-TR-2. We also plot a theoretical main-sequence isochrone derived by \cite{Baraffe1998}. As noted by \cite{RibasReview} and references therein, stars with masses greater than  $\sim$0.3~\msun\ seem to fall systematically above the theoretical line, with their radii being underestimated by $\sim$10\%. The two stars in UNSW-TR-2 appear to agree with this trend, although a more precise determination of their masses and radii is clearly required.

It is likely that one or both of our stars lies in the previously unpopulated region between 0.5 and 0.6~\msun, which increases their importance in constraining the mass-radius relation of low-mass stars. The apparent magnitude of our system ($V\simeq13.9$) is similar to that of the 0.6~\msun\ eclipsing binary GU Bootis ($V\simeq13.7$), for which \cite{GUBoo} derived the masses and radii to within the 2--3\% accuracy required for a detailed comparison with theory. With echelle spectroscopy and further photometry, the properties of UNSW-TR-2 could therefore be similarly constrained.

\subsection{UNSW-TR-2 as a target for a planet search}

Eclipsing binary systems with dwarf components are interesting targets for photometric planet searches. One main advantage is that the orbits of planets in such a system are likely to be near-coplanar with the binary, making transits more likely. Additionally, non-transiting planets may also be detected via the variations they induce in the timing of the binary eclipse minima \citep[e.g.][]{DoyleDeeg2004}. These motivated an extensive search around the M-dwarf binary CM~Dra \citep{Doyle2000}.

In the case of UNSW-TR-2, transits of a Jupiter-sized planet in front of one of the stars (when they are not eclipsing) would be 1.3\% deep. A planet in a co-planar orbit about the binary ($i=83.5$\dg) would have to be within $r_{\rm max}\simeq0.04$~AU of the centre of mass to transit. This is only 2.5 times the binary's semi-major axis, but according to simulations by \citet{HolmanWiegert1999}, even such a small orbit may be stable around an equal-mass binary with a circular orbit. Planets with more favourable orbital inclinations (relative to the line of sight) may also transit. For example, $i_p=85$\dg\ gives $r_{\rm max}=0.05$~AU, and $i_p=87$\dg\ gives $r_{\rm max}=0.09$~AU.

The approximate range of eclipse-timing variations is $\tau = qr/c$, where $q$ is the planet-to-binary mass ratio, $r$ is the planet's orbital radius, and $c$ is the speed of light \citep{DoyleDeeg2004}. For a Jupiter-mass planet ($q\simeq10^{-3}$) in a 1~AU orbit, this gives $\tau \simeq 0.5$~sec. The detection of planets via this method would therefore require second-precision timing of eclipses over timescales of years.

\section{Conclusions}

We present a preliminary analysis of a new double-lined eclipsing
binary system with K7~Ve components.  The system was discovered as
part of an ongoing transit search for extra-solar planets. The system
is important in that it is only the seventh low-mass binary known, and
because it allows precise determinations of the masses and radii of
these low-mass stars.  Using the initial photometric data, and
follow-up photometry and spectroscopy, we derive the physical
parameters of the system.  Our results are consistent with previous
work, confirming a departure in the mass-radius plane from the
theoretical models of \cite{Baraffe1998}.  Enhanced magnetic activity
--- due to the high rotational velocities of stars in close binaries
--- may be responsible for this discrepancy \citep{RibasReview}.

We find statistically significant evidence --- in fact the most
significant in any known low-mass eclipsing binary system --- for a
small but non-zero eccentricity in this new binary system, suggesting
the possibility of an additional companion.


\section*{Acknowledgments}

The authors wish to thank A.~Phillips for continuous maintenance and upgrade work on the APT, H.~Toyozumi and C.~Blake for help with the follow-up observations, and M.~Irwin for assistance with developing the APT data reduction pipeline.
 We are also grateful for valuable comments from M.~L{\'o}pez-Morales, B.~Carter, C.~Maceroni, T.~Marsh and P.~Maxted, and for telescope time from the Mount Stromlo and Siding Spring Observatories TAC. JLC is supported by an Australian Postgraduate Research Award. We thank Sun Microsystems for their generous donation of a workstation.


\bibliographystyle{mn2e}
\bibliography{tr2paper}

\bsp

\label{lastpage}

\end{document}